\documentclass[aps,prd,preprint,superscriptaddress,nofootinbib,showpacs]{revtex4}

\usepackage{amsfonts}
\usepackage{mathrsfs}
\usepackage{amssymb}
\usepackage{amsmath}
\usepackage{graphicx,epsfig}
\usepackage{pstricks}
\usepackage{slashed}
\def\fo{\hbox{{1}\kern-.25em\hbox{l}}}
\input{epsf}
 \normalsize

\def\be{\begin{equation}}
\def\ee{\end{equation}}
\def\bea{\begin{eqnarray}}
\def\eea{\end{eqnarray}}

\def\non{\nonumber}
\def\lsim{\raise0.3ex\hbox{$\;<$\kern-0.75em\raise-1.1ex\hbox{$\sim\;$}}}
\def\gsim{\raise0.3ex\hbox{$\;>$\kern-0.75em\raise-1.1ex\hbox{$\sim\;$}}}

\def\ie{{\it i.e.}}
\begin{document}
\begin{flushright}
HIP-2011-14/TH \\
OSU-HEP-11-07
\end{flushright}

\title{Radion Flavor Violation in Warped Extra Dimension}
\vskip 0.2cm
\author{K. Huitu$^1$, S. Khalil$^{2,3}$, A. Moursy$^2$, S.K. Rai$^4$ and A. Sabanci}
\affiliation{$^1$ Department of Physics, and Helsinki Institute of
Physics, P.O.Box 64, FIN-00014 University of Helsinki, Finland\\
$^2$ Centre for Theoretical Physics, The British University in
Egypt, El Sherouk City, 11837, Egypt\\
$^3$Department of Mathematics, Ain Shams University, Faculty of
Science, Cairo, 11566, Egypt.\\
$^4$Department of Physics, and Oklahoma Center for High Energy Physics,
Oklahoma State University Stillwater, OK 74078, USA}

\date{\today }

\begin{abstract}
We analyze the flavor violation in warped extra dimension due to
radion mediation. We show that $\Delta S=2$ and $\Delta B=2$
flavor violating processes impose stringent constraints on radion
mass, $m_\phi$ and the scale $\Lambda_\phi$. In particular, for
$\Lambda_\phi \sim {\cal O}(1)$ TeV, $B_d^0 -\bar{B}^0_d$ implies
that $m_\phi \gsim 65$ GeV. We also study radion contributions to
lepton flavor violating processes: $\tau \to (e,\mu) \phi$, $\tau
\to e\mu^+\mu^-$ and $B\to l_i l_j$. We show that $BR(B_s \to
\mu^+ \mu^-)$ can be of order $10^{-8}$, which is reachable at the
LHCb. The radion search at LHC, through the flavor violation
decays into $\tau \mu$ or top-charm quarks, is also considered.

\end{abstract}
\maketitle

\section{Introduction} \label{sec:1}%
Extra dimensions have been proposed as an alternative way to
address the origin of the large scale discrepancy between Planck
scale and electroweak scale, known as hierarchy problem
\cite{ADD}. The warped extra dimension is one of the interesting
possibilities for a geometrical way to look at this problem. In the
Randall-Sundrum (RS) model with two branes \cite{Randall} the
electroweak scale is exponentially suppressed and a large
hierarchy between the Planck scale and the TeV scale is obtained.
The original RS model is based on the assumption that the Standard Model
(SM) fields are localized to one of the boundaries and gravity
only is allowed to propagate in the bulk. In this scenario, the
non-renormalizable operators in the 4-dimensional effective theory
are only TeV-scale suppressed. This would lead to rapid proton
decay and unacceptable flavor violation. If the SM fermions are
assumed to be propagating in the bulk one may be able to overcome
these problems and also explain the fermion mass hierarchy
\cite{Pomarol,Shafi}. However, it was shown that in this case flavor
changing neutral currents impose strong constraints on the
$4$-dimensional scale \cite{Shafi}.

The radius of extra dimension in the RS model is assumed to be fixed
by a given constant. Goldberger and Wise \cite{Goldberger}
proposed  a mechanism to understand the possible mechanism for
radius stabilization. It was shown that by adding a scalar field
to the bulk, a potential for the radion field is obtained that
dynamically generates a vacuum expectation value (VEV) of the
radion. This VEV, which is related to the size of the extra dimension,
can be naturally of the order of TeV. The radion field $\phi(x)$
arises as the pseudo-Goldstone boson associated with translation
symmetry breaking after stabilizing the extra dimension. In this
case, the radion mass is given by \cite{Goldberger}%
\bea%
m_{\phi}^2 = \frac{k^2 v_v^2}{3 M_5^3} \epsilon^2 e^{-2 k r_c \pi},%
\eea%
where $M_5$ is the 5-dimensional Planck scale, $k \simeq M_{Pl}$, $k
r_c \simeq 12$, $\epsilon \ll 1$. Therefore, the mass of radion is
typically of the order of a few GeV's. Hence, it may be the lightest new
(non-SM) particle in this type of model with warped geometry.

The radion phenomenology has been discussed in several papers
\cite{Csaki,radionpheno,Cheung:2000rw,radpheno} and recently
flavor changing neutral currents mediated by radion field, like $t
\to \phi c$  and $\epsilon_K$, have been considered \cite{Azatov}.
These analyses showed that the top decay does not impose any
significant constraint on the stabilization scale
$\Lambda_{\phi}$, but the CP violating parameter $\epsilon_K$ may
strongly constrain it. However, $\epsilon_K$ depends on the CP
violating phases assumed in the Yukawa matrices, therefore it
cannot be used to give a model independent constraint on
$\Lambda_{\phi}$. Our goal in this paper is to pursue this study
and consider possible constrains due the experimental bounds of
$\Delta S=2$ and $\Delta B =2$ processes. In addition, we consider
radion contributions to lepton flavor violating processes like
$\tau \to (e, \mu) \phi$ and $\tau \to e\mu^+\mu^-$, in addition
to $B\to l_i l_j$. We show that although the radion effects
enhance the amplitudes of these process, their branching ratios
remain below the current experimental limits. We also analyze the
search for radion at LHC. In particular, we focus on the flavor
violation decay of radion to $\tau \mu$ or to top-charm quarks.

The paper is organized as follows. In Section \ref{sec:2}, we briefly review
the radion interactions with the SM model fermions propagating in
the 5D bulk while the Higgs is localized on the TeV brane. We
emphasize the radion flavor violating couplings with the SM
fermions. Sections \ref{sec:3} and \ref{sec:4} are devoted for analyzing the radion
contributions to $\Delta S=2$ and $\Delta B =2$ transitions and
the constraints imposed on the scale $\Lambda_\phi$ and radion
mass. It turns out that the $B_d-\bar{B}_d$ mixing gives the
strongest bounds on $\Lambda_\phi$ and $m_\phi$. In Section \ref{sec:5} we
study the radion contribution to the decays $ B_s \to l_i l_j$.
The effects of the radion mediation in lepton flavor violating
processes like $\tau \to (e, \mu) \phi$ and $\tau \to e\mu^+\mu^-$
are described in Section \ref{sec:6}. The radion search at the LHC is
discussed in Section \ref{sec:7}. Finally, we give our conclusions in
Section \ref{sec:8}.

\section{Radion interactions with the SM fermions} \label{sec:2}
We consider the following $5D$ AdS space-time \cite{Csaki}:
\begin{equation}
ds^2 = \left(\frac{R}{z}\right)^2\left( e^{-2F} \eta_{\mu\nu}
dx^\mu dx^\nu  - (1+2F)^2 dz^2 \right),
\end{equation}
where $z$ refers to the conformally flat AdS background with $R'<
z< R$. $R$ is the AdS curvature and is given by $R= 1/k
\simeq 1/M_{Pl}$ while $R'\simeq 1/$TeV. The scalar function
$F(x,z)$ corresponds to the radion fluctuation around the
stabilized radius. From $5D$ Einstein's equations one can show
that the metric perturbation
$F(x,z)$ is given by%
\be%
F(x,z)=\frac{\phi(x)}{\Lambda_{\phi}}\,\left(\frac{z}{R'}\right)^2\ ,
\ee %
where $\Lambda_{\phi} \equiv \sqrt{6}/R'$. Therefore the square root
of the $5D$ metric determinant  $\sqrt{g}$ is
given at linear order on $F$, by %
\begin{equation} %
\sqrt{g}\approx \left(\frac{R}{z}\right)^5
\left(1-2\frac{\phi(x)}{\Lambda_{\phi}}\,\left(\frac{z}{R'}\right)^2\right)\ .%
\end{equation}
The 5D action for bulk fermions can be written as:
\bea%
S_f &=& \int d^4x dz \sqrt{g} \left[ {i \over 2} \left({\bar{\cal
Q}_i} \Gamma^A {\cal D}_A {\cal Q}_i - {\cal D}_A \bar{ {\cal
Q}_i} \Gamma^A {\cal Q}_i\right) + {c_{q_i} \over R} \bar{ \cal
Q}_i {\cal Q}_i + \left({\cal Q} \rightarrow {\cal U}, {\cal
D}\right) \right. \nonumber\\
&&+\left. \left(Y^u_{ij}\sqrt{R}\ \bar{{\cal Q}_i} {\cal H} {\cal
U}_j + Y^d_{ij}\sqrt{R}\ \bar{{\cal Q}_i} {\cal H}
{\cal D}_j  + h.c.\right) \right], %
\label{fermionaction} %
\eea%
where $\Gamma$ matrices are given by $\Gamma^A=\gamma^{a}
e^A_a$,$a=0,1,2,3,5$ stands for $5D$ Lorenz indices. $\gamma^a$
are the ordinary $\gamma$-matrices with $\gamma^5= i~ {\rm diag}
(1_2,-1_2)$. Here the $5D$ fermion mass is given in terms of the
scale $R$ and the bulk parameter $c_f$. We assume that the Higgs
field is localized on the TeV-brane, i.e ${\cal H}(x,z)= H(x)~
\delta(z-R')$. ${\cal Q}_i$, ${\cal U}_i$, and ${\cal D}_i$ are
the $5D$ fermions, with flavor indices $i,j=1,2,3$, which contain
the $4D$ SM $SU(2)_L$ doublet and singlet fermions, respectively.
They can be written in two component spinor notation as follows:
\bea {\cal Q}_i=\left(\begin{array}{c}{\cal Q}^i_L\\
\bar{\cal Q}^i_R\end{array}\right)~, \hspace{0.5 cm}
{\cal U}_i=\left(\begin{array}{c}{\cal U}^i_L\\
\bar{\cal U}^i_R\end{array}\right)~, \hspace{0.5 cm} {\rm and}
\hspace{0.5 cm} {\cal D}_i=\left(\begin{array}{c}{\cal D}^i_L\\
\bar{\cal D}^i_R\end{array}\right)\  .\eea

The Kaluza-Klein (KK) decomposition for the $5D$ bulk fields is,
as usual, given by %
\bea %
{\cal Q}_{L,R}(x,z) &=& \sum_n Q^n_{L,R}(z)~ q^n_{L,R}(x)%
\eea %
with similar expressions for ${\cal U}_{L,R}$ and ${\cal
D}_{L,R}$. The zero modes $q_L(x)$, $u_R(x)$ and $d_R(x)$
define the $4D$ SM fermions that satisfy the Dirac equations%
\bea %
&&- i \bar{\sigma}^{\mu} \partial_{\mu} q^i_L + m^u_{ij}
\bar{u}^j_R =0,\\
&&- i \bar{\sigma}^{\mu} \partial_{\mu} q^i_L + m^d_{ij}
\bar{d}^j_R =0, %
\eea %
where $m^{q}_{ij}$ are the mass matrices for up and down quarks
which generally are not diagonal in flavor space. Also $m^{q}_{ij}$ is
not simply the induced mass on the TeV brane, given by Higgs VEV and the
effective Yukawa coupling weighted by zero mode profiles. $m^{q}_{ij}$
is the mass eigenvalue that emerges from the solution of the
coupled bulk equations of motion, taking into account the Higgs
interactions. In general, the physical mass receives corrections
from the reaction of the wave-functions to the brane where the
Higgs is localized \cite{Csaki},
\begin{equation}
\left( {m^{q}}_{ij}\right)^2 = (M_D)^2_{ij} \frac{ (1-2c^i_L) (1+2c^j_R)}{\left( 1 -
\lambda^{1-2c^i_L} \right) \left( 1- \lambda^{1+2c^j_R} \right)
}\ \ , \label{eq:fermmass}
\end{equation}
where $\lambda \equiv R/R'$ and $M_D$ is the localized Dirac mass,
{\it i.e.}~induced mass on the brane through the Higgs VEV. Similar
expression for charged lepton masses can also be obtained. The
boundary conditions
are usually chosen such that %
\bea %
{\cal Q}_R|_{z=R,R'}={\cal U}_L|_{z=R,R'}={\cal
D}_L|_{z=R,R'}=0.%
\eea %
These conditions allow the $SU(2)_L$ doublet (left-handed state)
and singlet (right-handed state) only to have zero modes.
Moreover, due to the arising discontinuities, one should impose the
following conditions as well \cite{Csaki}:%
\bea %
&&{\cal Q}_R|_{R^{\prime\,-}} = -M^u_D R^\prime \, {\cal
U}_R|_{R^{\prime\, -}}, \\
&& {\cal U}_L|_{R^{\prime\,-}} = -M^u_D R^\prime \, {\cal
Q}_L|_{R^{\prime\, -}}.
 \eea
Similar boundary conditions are applied for ${\cal D}_{L,R}$. The
bulk wave functions can be found in Ref.\cite{Csaki}. For the zero
modes, with the approximation $m R^\prime \ll 1$, the associated
wave functions can be written as
\begin{eqnarray}
Q^0_{L}(z) &=& z^{2} A^q_L \left[ \frac{m
z^{c^q_{L}+1}}{2c^q_{L}+1} + \frac{z^{-c^q_{L}} R^{2 c^q_L -1}}{m}
\left(1-\frac{m^2 z^2}{2-4c^q_{L}}\right) \right],
\nonumber \\
Q^0_{R}(z) &=& z^{2} A^q_R \left[ z^{c^q_R} \left(1-\frac{m^2
z^2}{2+4 c^q_{R}}\right) + m^2 \frac{R^{1+2 c^q_R}}{1+2 c^q_R}~
\frac{z^{1-c^q_{R}}}{1-2 c^q_R} \right].
\end{eqnarray}
The parameters $A^q_{L,R}$ can be determined from the
normalization conditions: %
\bea%
\int dz \left(\frac{R}{z}\right)^4 \left\vert Q^0_{L,R} \right\vert^2 =1.%
\eea%
Similar expressions can be obtained for $U^0_{L,R}$ and
$D^0_{L,R}$. In this respect, the general expression for the
Lagrangian of
radion interaction with SM fermions is given by \cite{Csaki}%
\bea%
{\cal{L_\phi}}=\frac{\phi(x)}{\Lambda_{\phi}} (q^i_L \bar{u}^j_R + \bar{q}^i_L
u^j_R)\!\!\!\! &\times& \!\!\!\!\int dz \left(\frac{R}{z}\right)^2
\frac{R^2}{R'^2}
\left[ -\frac{m^u_{ij}}{2} \left( (Q^i_L)^2 + (Q^i_R)^2\right) \right.\nonumber\\
\!\!\!\!&+&\!\!\!\! \left. 2 \left(Q^i_L (Q^i_R)'-(Q^i_L)' Q^j_R +
\frac{c^{q_i}_L}{z} Q^i_L Q^i_R\right) + \left(
Q^i_{L,R}\rightarrow U^j_{L,R}, D^j_{L,R} \right)\right].~~~
\label{radionferm} \eea %
In addition
to the approximation  $m R' \ll 1$, for light fermions which are usually assumed to be localized near
the Planck brane {\it i.e}, $c_L>1/2$ and $c_R<-1/2$ one finds that the associated
radion couplings take the following simple form:
\begin{equation}
\frac{m^{u,d}_{ij}}{\Lambda_{\phi}}  (c^{q_i}_L-c^{u_j,d_j}_R).
 \label{lightfermcoupl}
\end{equation}
Transforming to mass eigenstate via the unitary matrices, $V^d_L$
(obtained by left-handed quark rotation) and $V^d_R$ (obtained by
right-handed quark rotation) will diagonalize the down mass
matrix. In this basis, the radion couplings with down quarks
$Y_{\phi d_id_j}$ are non-universal and are given by %
\bea Y_{\phi d_i d_j} = (V^d_L)^T_{ik}~ .
\frac{m^d_{kl}}{\Lambda_{\phi}}
(c^{q_k}_L-c^{d_l}_R)~.~ (V^d_R)_{lj}.%
\label{lightfermcoup2}\eea%
 It is clear that this
flavor violation can be mediated at tree level by the radion
propagation, which might be quite dangerous and lead to strong
bounds on the stabilization scale $\Lambda_{\phi}$.

Note that the bulk mass parameters $c^{u,d}_{L,R}$ and $5D$ Yukawa
couplings are free parameters to be fixed by the observable masses
and mixing. Therefore, in this class of models the number of free
parameters is larger than the number of the quark masses and
mixings. In our analysis, as an example, we consider the following
values of $c^{u,d}_{L,R}$ that lead to consistent quark masses at
the weak scale\footnote[1]{We modify the model in \cite{Shafi} by imposing the conditions for c-parameters of light fermions {\it i.e} $c_L>1/2$ and $c_R<-1/2$.}

\begin{eqnarray}
c_{q1}^L=0.72,~~ & c_{d1}^R=-0.57,~~ & c_{u1}^R=-0.63  \nonumber \\
c_{q2}^L=0.60,~~ & c_{d2}^R=-0.57,~~ & c_{u2}^R=-0.30  \nonumber \\
c_{q3}^L=0.35,~~ & c_{d3}^R=-0.60,~~ & c_{u3}^R=-0.10
\label{qcparameter}
\end{eqnarray}
We also fix the Dirac mass ${(M_D)}_{ij}= v l_{ij}$, where
$l_{ij}$ are dimensionless quantities of order unity obtained from
the 5D Yukawa couplings and $v$ is taken to be the SM VEV, namely
\begin{eqnarray}
l^d_{ij}=\left(\begin{array}{ccc}
0.50 & -2.00 & -2.00 \\
1.48 & -0.90 & 2.00 \\
0.52 & -0.50 & 0.70
\end{array}\right), \quad
l^u_{ij}=\left(\begin{array}{ccc}
0.80 & -1.90 & -2.00\\
1.23 & 1.20 & -1.04 \\
1.85 & 1.66 & -0.80
\end{array}\right).
\label{lijparameters}
\end{eqnarray}
In general, these parameters are complex. However, the
corresponding phases may lead to a large contribution to the CP
violating processes (as $\epsilon_K$), which is inconsistent with
the SM expectations. Therefore, these phases are typically
constrained and set to zero unless one assumes a specific texture
of flavor that suppresses both CP conserving and CP violating flavor
changing effects as in Ref.\cite{Delaunay:2010dw}

Using these parameters, one obtains the following quark masses:
\begin{eqnarray}\label{quarkmasses}
&m_d=2.7~{\rm MeV}, \quad &m_u=1.66~{\rm MeV}, \nonumber \\
&m_s=47~{\rm MeV}, \quad & m_c=1.2~{\rm GeV} , \nonumber \\
&m_b=3.524~{\rm GeV}, \quad & m_t=171.25~{\rm GeV}
\end{eqnarray}
and the CKM matrix is given by %
\bea %
V_{CKM}=\left(%
\begin{array}{ccc}
  0.972061 & 0.23468 & 0.00473366 \\
  0.234685 & 0.971301 & 0.0386956 \\
  0.00448328 & 0.0387254 & 0.99924 \\
\end{array}%
\right)\eea

Corresponding equations can be written for leptons as well, and thus
acceptable lepton masses can be derived from
Eq.~(\ref{eq:fermmass}) where $q$ has been replaced by $l$, using the following parameters%
\begin{eqnarray}\label{Lcparameter3}%
\begin{array}{cc}
c_{l1}^L=0.705,~~ & c_{R}^{e}=-0.52~~    \\
c_{l2}^L=0.655,~~ & c_{R}^{\mu}=-0.53~~  \\
c_{l3}^L=0.550,~~ & c_{R}^{\tau}=-0.585~~
\end{array},\;\;\;\;l_{ij}^l=\left(
             \begin{array}{ccc}
               -1.37& 1.19 & 1.04 \\
               1.10 & 1.10 & 0.90 \\
               -0.80 & 0.90 & 0.90 \\
             \end{array}\right).
\end{eqnarray}
In this case, one finds $m_e= 0.511$ MeV, $m_{\mu}= 0.106$
GeV, and $m_{\tau}= 1.777$ GeV. Also the Yukawa couplings of
radion-fermion-antifermion can be approximately written, in terms
of the scale $\Lambda_\phi$, as
\bea%
Y_u&=&{1\,{\rm GeV}\over \Lambda_\phi}\left(%
\begin{array}{ccc}
  0.00224209 & 0.0209379 & 0.146558 \\
  -0.00311323 & 0.91945 & 0.00224209 \\
  -0.445271 & 15.937 & 100.47 \\
\end{array}%
\right)~,\\
Y_d&=&{1\,{\rm GeV}\over \Lambda_\phi}\left(%
\begin{array}{ccc}
  0.00347354 & -0.000530211 & -0.00618497 \\
  -0.000427834 & 0.055131 & -0.0546559 \\
  0.0347305 & -0.0138639 & 3.26096 \\
\end{array}%
\right)~,\eea
and
\bea %
Y_e={1\,{\rm GeV}\over \Lambda_\phi}\left(
\begin{array}{ccc}
  0.000657813 & -0.000167775 & -0.00391232 \\
  -0.000800194 & 0.124867 & 0.00101446 \\
  0.0151318 & -0.0109371 & 1.91215 \\
\end{array} \right).%
\label{yukawaE}%
\eea
The $c$- and $l$- parameters found here are obviously not unique and
one may wonder how general our results are using these sets.
To study that, we have generated another parameter set both for quarks and leptons.
Although the $c$- and $l$-values in the new sets are clearly different from the ones shown here
and used in the analyses of the latter sections, it turns out that the results
remain qualitatively the same, and quantitatively change only little.

\section{Radion contribution to $\Delta S=2$ transitions} \label{sec:3}
We start our analysis for radion flavor violation by considering
the radion contribution to $\Delta S=2$ processes, where $S$
refers to the $s$-quark number, in particular to $K^0 -
\bar{K}^0$. Generically, the $K_L-K_S$ mass difference $\Delta M_K$ is defined as %
\bea%
\Delta M_K = 2 \vert \langle K \vert H_{\rm eff}^{\Delta S=2}
\vert
\bar{K} \rangle \vert, %
\eea %
where $H_{\rm eff}^{\Delta S=2}$ is the effective Hamiltonian for
$\Delta S=2$ transition. With radion contribution to the
off-diagonal entry in the $K$-meson, the mass matrix ${\cal
M}_{12}(K) = \langle K \vert H_{\rm eff}^{\Delta S=2} \vert
\bar{K} \rangle$ is
given by%
\bea%
{\cal M}_{12}(K) = {\cal M}_{12}^{\rm SM} (K) + {\cal
M}_{12}^{\rm rad}(K).%
\eea%
Here ${\cal M}_{12}^{SM}(K)$ is the SM contribution and is given by%
\bea%
\label{KKbarSM} %
{\cal M}_{12}^{\rm SM}(K) = \frac{G_F^2}{12 \pi^2}\hat{B}_K f_K^2
M_K M_W^2 \left(\eta_1(\lambda^*_c)^2
S_0(x_c)+\eta_2(\lambda^*_t)^2 S_0(x_t)+
2\eta_2(\lambda^*_t)(\lambda^*_c) S_0(x_c,x_t) \right)%
\eea where $\lambda_i=V^*_{is}V_{id}$ and other parameters and
loop functions which appear in the above equation can be found in
Ref.~\cite{KKSM}. The SM expectation for $\Delta M_K$ is given by %
\be%
\Delta M_K^{\rm{SM}} = 2.7018 \times 10^{-15} \textrm{GeV}\ ,%
\ee %
which lies in the ballpark of the measured value \cite{pdg1}:%
\bea %
\Delta M_K^{\textrm{exp}}=3.483\pm 0.006\times 10^{-15}
\textrm{GeV}\ .%
\eea%
However, a precise prediction cannot be made due to the hadronic
and CKM uncertainties.

%
\begin{figure}[t]
\begin{center}
\hspace*{-7mm}
\includegraphics[width=6cm,height=3cm]{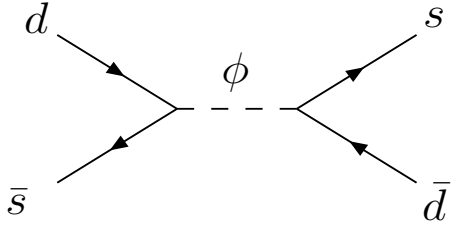}\\
\caption{{\small Radion contributions to the $K^0-\bar{K}^0$
mixing.}} \label{fig1}
\end{center}
\end{figure}

Unlike the SM, the radion contribution to the $K^0 -\bar{K}^0$
mixing is at the tree level, as shown in Fig.1. The corresponding
effective Hamiltonian is given by %
\bea %
H_{\rm eff}^{\rm rad} (\Delta S=2) = \sum_{i=1,2} \left( C_i Q_i +
\tilde{C}_i
\tilde{Q}_i\right)~, %
\eea %
where $C_i$, $\tilde{C}_i$, $Q_i$ and $\tilde{Q}_i$ are the Wilson
coefficients and operators with %
\bea \label{operators} Q_1 &=& (\bar{s}_L d_R) (\bar{s}_L
d_R),~~~~~~~~~~~~~~Q_2 = (\bar{s}_L
d_R) (\bar{s}_R d_L),\nonumber \\
C_1 &=&-\frac{Y_{\phi s_Ld_R}^2}{M_K^2-m_{\phi}^2 + i m_\phi
\Gamma_\phi} ,~~~~~~~~~~~~~~~ C_2=-\frac{Y_{\phi s_Ld_R}Y_{\phi
s_Rd_L}}{M_K^2-m_{\phi}^2 + i m_\phi \Gamma_\phi}.%
\eea%
The operators $\tilde{Q}_i$ and Wilson coefficients $\tilde{C}_i$
are obtained from $Q_i$ and $C_i$ by exchanging $L \leftrightarrow
R$. Note that for $m_\phi \gg M_K$, the Wilson coefficients are
given by $\sim Y^2/M_\phi^2$. For $m_\phi < M_K$, if we assume
that the momentum transfer in the four-fermion operator is around $M_K$
the coefficients are $\sim Y^2/M_K^2$, which is a consistent
approximation since for light radion the $K-\bar{K}$ transition
occurs through the decay of $K$ into $\phi$ and $X_d$.

The mass of the radion is in the range of a few GeVs when the
external momenta are neglected. The matrix elements of the
operators $Q_i$ between $K$ mesons in the Vacuum Insertion
Approximation (VIA) are given by \cite{Bparameter2}:
\begin{eqnarray}
        \langle \bar{K}^{0} \vert Q_{1}
        \vert K^{0} \rangle_{\rm VIA}  & = & -\frac{5}{24}
        \left(\frac{M_K}{m_{s}+m_{d}}\right)^{2}M_Kf_{K}^{2}\; ,
        \nonumber \\
        \langle \bar{K}^{0} \vert Q_{2}
        \vert K^{0} \rangle_{\rm VIA} & = & \left[\frac{1}{24} +
        \frac{1}{4}
        \left(\frac{M_K}{m_{s}+m_{d}}\right)^{2}\right]M_Kf_{K}^{2}\; ,
        \label{me}
\end{eqnarray}
where $m_{s}$ and $m_{d}$ are the masses of $s$ and $d$ quarks,
respectively. In the case of the renormalized operators, we define
the $B$-parameters as
 \bea \langle \bar K^{0} \vert \hat Q_{1}
(\mu) \vert K^{0} \rangle &=& -\frac{5}{24} \left( \frac{ M_K }{
m_{s}(\mu) + m_d(\mu) }\right)^{2}
M_K f_{K}^{2} B_{1}(\mu) ,\nonumber  \\
\langle \bar K^{0} \vert \hat Q_{2} (\mu) \vert K^{0} \rangle &=&
\frac{1}{4} \left( \frac{ M_K }{ m_{s}(\mu) + m_d(\mu)
}\right)^{2} M_K f_{K}^{2} B_{2}(\mu)\label{eq:bpars} \ , \eea
where $\hat Q_{i}(\mu)$ denotes the operators renormalized at the
scale $\mu$. For the scale $\mu=2$ GeV, $B_{1}(\mu)$ and
$B_{2}(\mu)$ are given by \cite{Bparameter2}: \bea
  \label{eq:fres}
B_{1}(\mu) &=& 0.66(4)  \nonumber \\
B_{2}(\mu) &=& 1.03(6)
 \eea
\begin{figure}[!ht]
\includegraphics[width=13cm,height=8.5cm]{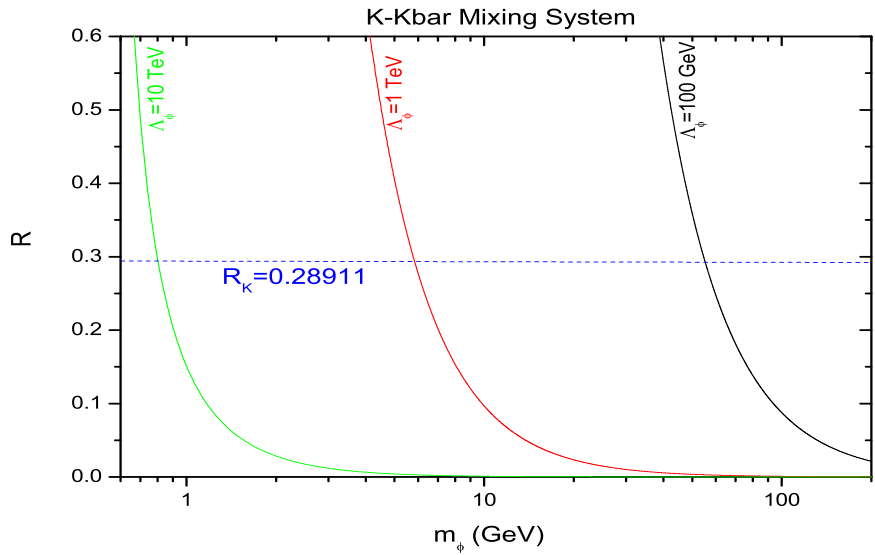}
\caption{{\small The ratio $R_K = \Big \vert \frac{{\cal
M}_{12}^{rad}(K)}{{\cal M}_{12}^{SM}(K)}\Big\vert$ as function of
radion mass $m_\phi$ for scale $\Lambda_\phi = 0.1,1, 10$ TeV.}}
\label{fig2}
\end{figure}

Using Eq.(\ref{lightfermcoup2}) and the values of the
$c$-parameters in Eq.~(\ref{qcparameter}), we can compute the
values of the Wilson coefficients at the scale of the radion mass.
Since the $K-$decay occurs at 2 GeV, we should run the Wilson
coefficients from the scale of the radion mass to the scale of 2
GeV, considering all thresholds, using the following general RGE
equations \cite{runningwilson} that runs the Wilson coefficients
from scale M to another scale $\mu$
\bea\label{runningcoefficients}
C_i(\mu)=\big[1+{\alpha_s(\mu)-\alpha_s(M)\over4\pi}J_f\big]\big[{\alpha_s(M)\over
\alpha_s(\mu)}\big]^{d_f}C_i(M) , \eea
where
\bea d_f= {\gamma^{(0)}\over 2 \beta_0},\hspace{0.5cm} J_f={
d_f\over \beta_0}\beta_1-{\gamma^{(1)}\over 2
\beta_0}\nonumber,\eea
and
\bea\gamma^{(0)}&=&6\frac{N- 1}{ N} , \hspace{0.5cm}
\gamma^{(1)}={\frac{N- 1}{2N}}\big[-21+{57\over N} -{19\over 3}N +
{4\over3}f\big] ,\nonumber \\
\beta_0&=&{(11N-2f)/3}, \hspace{0.5cm} \beta_1={34\over
3}N^2-{10\over 3}N f-{N^2-1\over N}f ,\eea
where $N$ is the number of colors and $f$ is the number of active
flavors. Also we run the produced masses, Eq.~\ref{quarkmasses},
from the weak scale to the scale of 2 GeV.

Using the above expressions, one can compute the radion
contribution to $\Delta M_K$. The experimental limits of $\Delta
M_K$ implies that $R_K = \Big \vert \frac{{\cal
M}_{12}^{rad}(K)}{{\cal M}_{12}^{SM}(K)}\Big\vert \leq 0.28911$,
which leads to an upper bound on the radion contribution. This
upper bound imposes a stringent constraint on the scale
$\Lambda_\phi$ and the radion mass $m_\phi$.
In Fig.2 we show the constraint on the radion mass $m_{\phi}$, due
to the $K-\bar{K}$ mixing system, for three values of the scale
$\Lambda_{\phi}: 0.1, 1,$ and $10$ TeV. As can be seen from this
figure, a very light radion ${\cal O}$($1$) GeV can be allowed if
$\Lambda_{\phi}$ of order $10$ TeV. However $\Lambda_{\phi} \lsim
1$ TeV can be consistent with $\Delta M_K$ experiment bound if
$m_\phi\geq {\cal O}(5)$ GeV.

In order to study the sensitivity of these bounds on $m_\phi$ and
$\Lambda_\phi$ to the values of the bulk mass
parameters $c$ and $5D$ Yukawa parameters $l$, we consider
another example of these parameters that produce the correct quark
masses and $V_{CKM}$ mixing matrix. Namely, the following set of
parameters is considered:
\begin{eqnarray}
c_{q1}^L=0.643,~~ & c_{d1}^R=-0.675,~~ & c_{u1}^R=-0.645  \nonumber \\
c_{q2}^L=0.583,~~ & c_{d2}^R=-0.630,~~ & c_{u2}^R=-0.630  \nonumber \\
c_{q3}^L=0.317,~~ & c_{d3}^R=-0.590,~~ & c_{u3}^R=~0.150
\label{qcparameter2}
\end{eqnarray}

\begin{eqnarray}
 l_{ij}^d=\left(
             \begin{array}{ccc}
              1.1& 0.49 & 0.88 \\
               0.86 & -0.96 & -0.53 \\
               0.99 & 1.1 & -1.20 \\
             \end{array}\right), ~~~
           l_{ij}^u=\left(\begin{array}{ccc}
           -0.44& 1.21 & -0.50 \\
           -0.91 & -0.24 & 1.22 \\
            0.40 & -1.15 & 0.99  \\
\end{array}\right).
\label{lijparameters2}
\end{eqnarray}
Although these parameters are different from the ones listed in
Eq.(\ref{qcparameter},\ref{lijparameters}), we find out that there
is no important difference between the two examples and the above
bounds derived on $m_\phi$ and $\Lambda_{\phi}$ remains intact.
\section{Radion contribution to $\Delta B=2$ transitions} \label{sec:4}
There are two neutral $B^0 -\bar{B}^0$ meson systems: $B_q^0
-\bar{B}_q^0$, with $q=d,s$. In this systems, the flavor
eigenstates are given by $B_q=(\bar{b}q)$ and
$\bar{B}_q=(b\bar{q})$. As in the $K^0 -\bar{K}^0$ system, the
mass difference between mass eigenstates $B_L^q$ and $B_H^q$ is
defined as %
\be%
\Delta M_{B_q} = M_{B_H}^q - M^q_{B_L} = 2 \vert {\cal
M}^q_{12}\vert = 2 \vert \langle B_q^0 \vert H_{\rm eff}^{\Delta B=2} \vert \bar{B}_q^0 \rangle \vert .%
\ee%
The experimental values for mass difference for $\Delta M^{\rm exp}_{B_d}$
and $\Delta M^{\rm exp}_{B_s}$ are given by \cite{pdg1}
\begin{eqnarray}%
\label{bbbarexperimental}%
\Delta M^{\rm exp}_{B_d} &=& (3.337\pm 0.033) \times 10^{-13} {\rm GeV}
\;,\\
\Delta M^{\rm exp}_{B_{s}} &=& (117.0 \pm 0.8 )\times10^{-13} {\rm GeV} \; .
\end{eqnarray}
The SM contribution for $\Delta M_{B_q}$ at NLO is given
by \cite{burasSM}
\begin{eqnarray}
\Delta M_{B_q}^{\rm SM}=\frac{G_F^2}{6\pi^2}\eta_B
m_{B_{q}}\hat{B}_{B_{q}}F^2_{B_q}M_W^2S_0(x_t)(V^*_{tq}V_{tb})^2
\end{eqnarray}
where $F_{B_q}$ is the $B_q$ meson decay constant for $q=d,s$ and
$\hat{B}_{B_q}$ is the renormalization-group invariant parameters \cite{Lellouch}.
One can show that the SM predictions for $\Delta M_{B_q}$ are given by %
\be%
\Delta M_{B_d}^{\rm SM} = 3.58187  \times 10^{-13} {\rm
GeV},~~~~~~~~~~ \Delta M_{B_s}^{\rm SM} = 104.19
\times 10^{-13} {\rm GeV}.%
\ee

The leading diagrams of radion contributions to the
effective Hamiltonian approach $H_{\rm eff}^{\Delta B_q=2}$ are
given by tree level diagrams similar to the diagram of $K^0
-\bar{K}^0$ mixing, with replacing $s$-quark by $b$-quark and
$d$-quark by $q$-quark. The induced effective Hamiltonian for
$\Delta B=2$ radion mediated process is given by %
\begin{eqnarray}
H_{\rm eff}^{\rm rad}(\Delta B_q=2) = \sum_{i=1,2} \left( C_i Q_i +
\tilde{C}_i \tilde{Q}_i\right)~,
\end{eqnarray}
where the operators $Q_i$ and the Wilson coefficients $C_i$ are given by%
\bea %
\label{BBbaroperators}%
Q_1&=& (\bar{b}_L q_R) (\bar{b}_L q_R),~~~~~~~~~~~ Q_2 = (\bar{b}_L q_R) (\bar{b}_R q_L),\nonumber \\
C_1&=& \frac{Y_{\phi b_L q_R}^2}{M_{B_q}^2-m_{\phi}^2+i m_\phi
\Gamma_\phi},~~~~~~~~~~~~~C_2=\frac{Y_{\phi b_L q_R}Y_{\phi b_R
q_L}}{M_{B_q}^2-m_{\phi}^2+i m_\phi
\Gamma_\phi}. %
\eea %
The operators $\tilde{Q}_i$ and the coefficients $\tilde{C}_i$ are
obtained from $Q_i$ and $C_i$ by exchanging $L \leftrightarrow R$. Here,
all the approximations on Wilson coefficients in the section (\ref{sec:3})
can be applied via replacing $M_K$ by $M_{B_q}$. Thus, the renormalized
hadronic matrix elements for radion mediated process can be found as \cite{becirevic}:
\begin{eqnarray}
\langle\bar{B}_q|\hat{Q}_1(\mu)|B_q\rangle&=&-\frac{5}{24}\left(\frac{m_{B_q}}
{m_b(\mu)+m_q(\mu)}\right)^2m_{B_q}F^2_{B_q} B_1(\mu) ,\nonumber\\
\langle\bar{B}_q|\hat{Q}_2(\mu)|B_q\rangle&=&\frac{1}{4}\left(\frac{m_{B_q}}
{m_b(\mu)+m_q(\mu)}\right)^2m_{B_q}F^2_{B_q} B_2(\mu).
\end{eqnarray}
Here we adopt the numerical values of $B_1$, $B_2$, $m_{B_q}$, and
$F_{B_q}$ as in \cite{becirevic}. Also after calculating the
Wilson coefficients at  the scale of the radion mass, we derive
the
\begin{figure}[!ht]
\includegraphics[height=6.0cm]{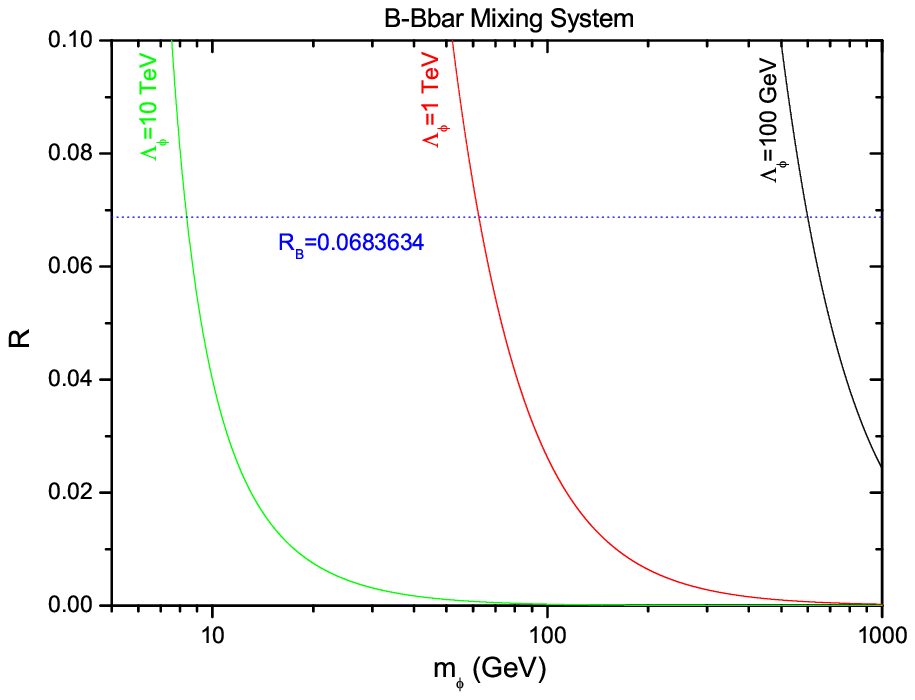}
\includegraphics[height=6.0cm]{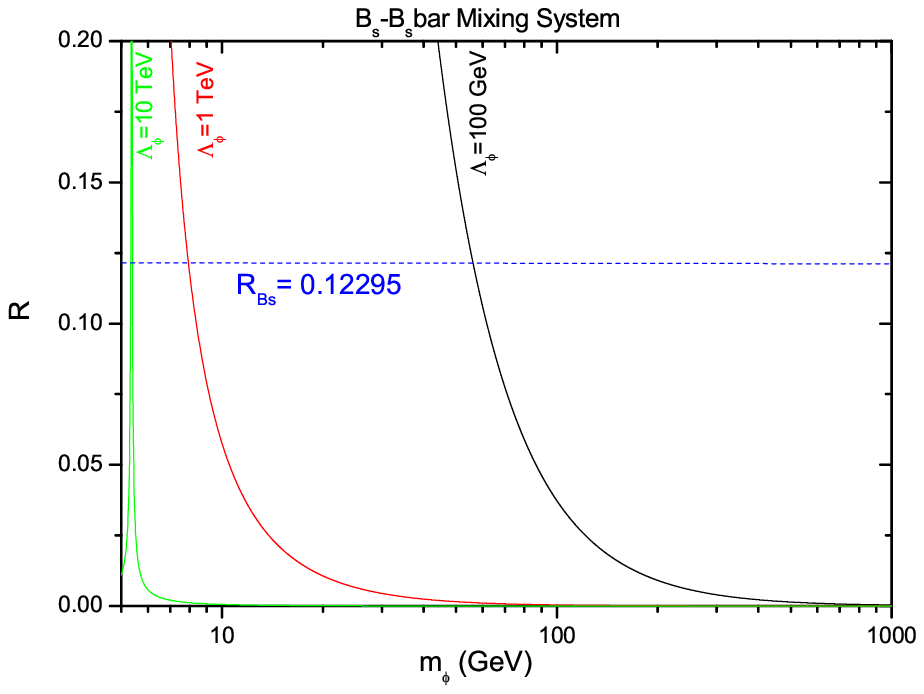}
\caption{{\small The constraints imposed on the radion mass
$m_\phi$, due to the radion contribution to ${B^0_d}-\bar{B}^0_d$
and ${B^0_s}-\bar{B}^0_s$ mixing systems, for $\Lambda_\phi
=0.1,1,10$ TeV.
 }}\label{BBbar}
\end{figure}
corresponding coefficients at $\mu=4.6$ GeV via Eq.
(\ref{runningcoefficients}). We consider the allowed upper bounds
on $\Delta M^{\rm exp}_{B_q}$ to derive new constraints on the radion free
parameters in our analysis: $\Lambda_\phi$ and $m_\phi$.

In Fig. \ref{BBbar} we present the constraints imposed on $m_\phi$
from the experimental results (using central values of the
results) for ${B^0_d}-\bar{B}^0_d$ and ${B^0_s}-\bar{B}^0_s$
 mixing systems, for $\Lambda_\phi =0.1,1,10$ TeV.
This figure shows that the experimental limits of $B_d^0
-\bar{B}_d^0$ gives more stringent constraints on $\Lambda_\phi -
m_\phi$ than the limits of $B_s^0 -\bar{B}_s^0$ and $K^0
-\bar{K}^0$. In this respect, it is clear that the processes of
$\Delta B=2$ and $\Delta S=2$ flavor violation play important role
in constraining the radion mass and it is no longer a free
parameter. For instance if we require that $\Lambda_\phi \sim
{\cal O}(1)$ TeV, which is favored by solving the hierarchy
problem, one finds that the radion mass has the following lower
bound: $m_\phi \gsim 65$ GeV. We have checked that these bounds
are not sensitive to the values of the bulk mass $c$-parameters
and $5D$ $l$-Yukawa parameters . We obtained very close limits on
$\Lambda_\phi$ and $m_\phi$ when we considered the example in
Eqs.(\ref{qcparameter2},\ref{lijparameters2}).

To our knowledge, it is the first time that such a lower bound on
radion mass is derived. Nevertheless, if a larger value of the
scale $\Lambda_\phi$ is considered, \ie, $\Lambda_\phi \gsim 10$
TeV, a smaller radion mass, $m_\phi \sim 10$ GeV, can be allowed.
As we will show below, a very light radion scenario is stringently
constrained by the lepton flavor violation decays $\ell_i \to
\ell_j \phi$.

\section{Radion contribution to leptonic $B$-decays} \label{sec:5}
We now consider the radion contribution to the leptonic
$B$-decays: $B_q \to l_i l_j$, where $q \equiv d,s$. In this class
of models with warped geometry, the $B$--meson decay into leptons
can be generated at tree level through radion exchange.
\begin{figure}[!ht]
\begin{center}
\hspace*{-7mm}
\includegraphics[width=6cm,height=3cm]{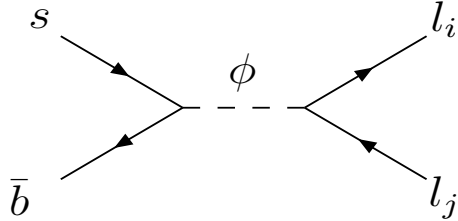}\\
\caption{{\small Radion contributions to the $B_s\rightarrow l_i
l_j$.}} \label{fig4}
\end{center}
\end{figure}
Note that, as in the quark sector, the radion couplings with the charged leptons are given by%
\bea%
Y_{\phi l_i l_j} = \frac{m^l_{ij}}{\Lambda_{\phi}} (c^{l_i}_L - c_R^{e_j}),%
\eea%
which transforms to the following expression in lepton mass basis:
\bea%
Y_{\phi l_i l_j} = (V^l_L)^T_{ik}
\frac{m^l_{kk'}}{\Lambda_{\phi}} (c^{l_k}_L - c_R^{e_{k'}})
(V^l_R)_{k'j}. \eea It is worth noting that these non-universal
couplings are obtained due to the mismatch between the
diagonalization of charged lepton mass matrix $m^l_{ij}$ and the
charged lepton-radion couplings $ \sim m^l_{ij} (c^{l_i}_L -
c_R^{e_j})$. The transition amplitude of this process is given by
\bea %
{\cal A}(\bar{B}_q \to l_i^- l_j^+)&=& \left[{1 \over
q^2-m_{\phi}^2}\right]\Big[Y_1 Y_3\bar{l}_i P_R
l_j\langle0|\bar{s}P_L b|B_q\rangle+Y_1Y_4\bar{l}_iP_R
l_j\langle0|\bar{s}P_R b|B_q\rangle \nonumber\\
&+&  Y_2Y_3\bar{l}_iP_L l_j\langle0|\bar{s}P_L b|B_q\rangle+ Y_2
Y_4\bar{l}_i P_L l_j\langle0|\bar{s}P_R b|B_q\rangle\Big],\eea
where the Yukawa couplings $Y_i, i=1,..,4$ are defined as%
\bea%
Y_1 &\equiv&  Y_{\phi l_{iL}^- l_{jR}^+},~~~~~~~~~~ Y_2\equiv
Y_{\phi l_{iR}^- l_{jL}^+},  \\
Y_3 &\equiv& Y_{\phi b_{L} q_{R}},~~~~~~~~~~ Y_4\equiv Y_{\phi
b_{R} q_{L}}.%
\eea%
The hadronic matrix elements are characterized by the decay
constant of the pseudoscalar meson $B_q$ and can be written as
\cite{Bobeth} %
\bea%
\label{matrix:element:btoll:II}%
\langle 0| {\bar{s}\gamma_5 b}|{B_s(p)}\rangle = - i
f_{B_s} \frac{M_{B_s}^2}{m_b+m_s}. %
\eea
\begin{figure}[t]
\includegraphics[height=6.5cm]{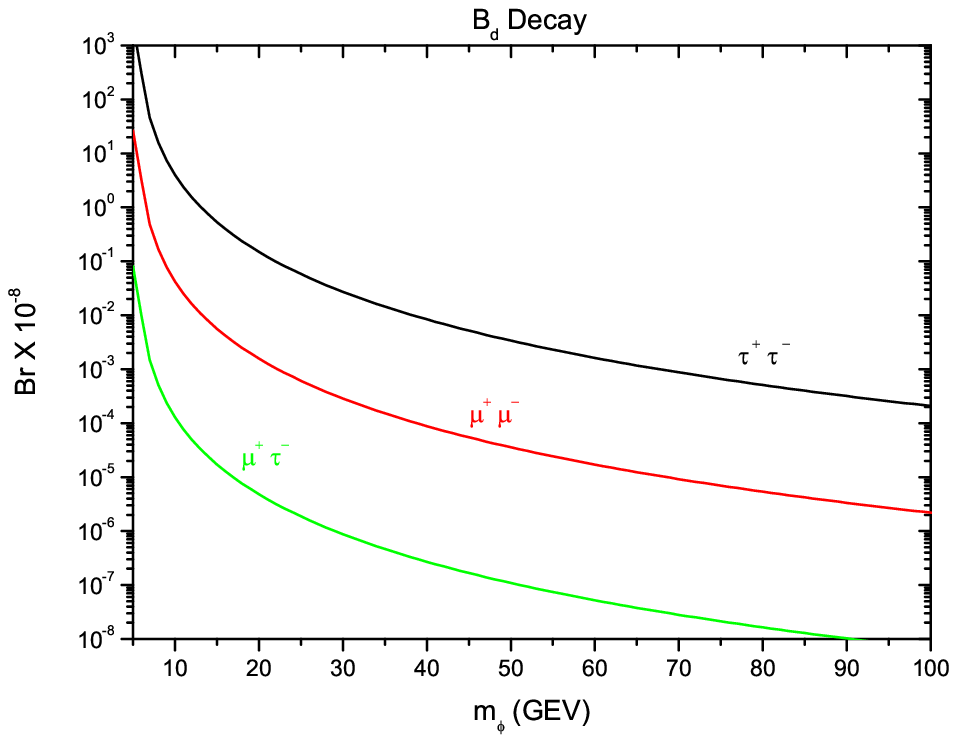}
\includegraphics[height=6.5cm]{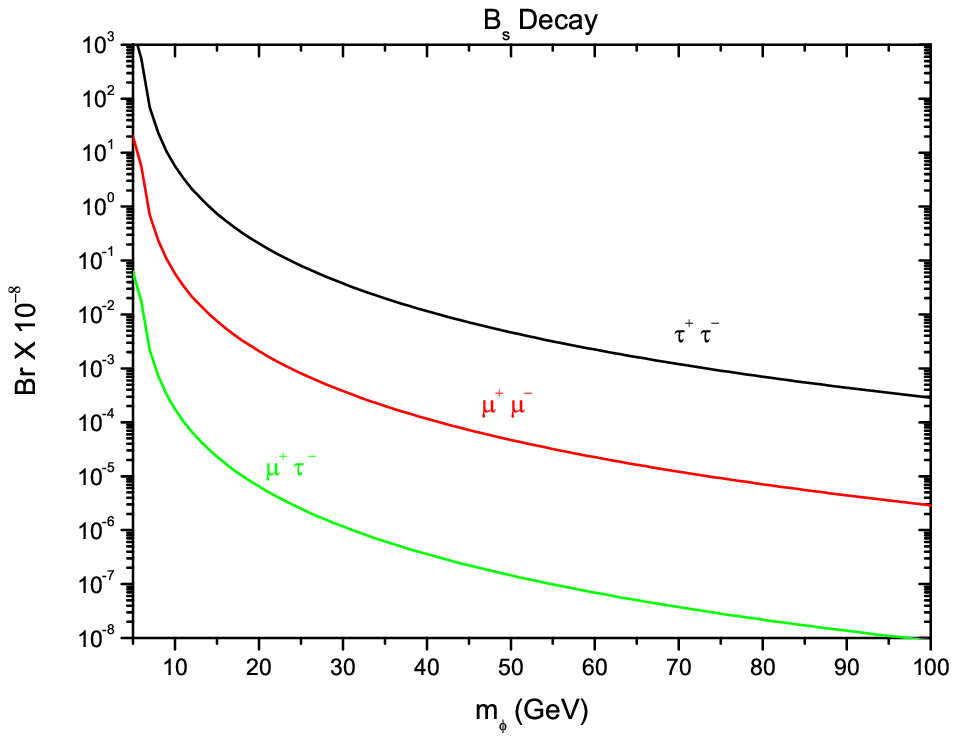}
\caption{{\small The branching ratios of the decays of $B_s$ and
$B_d$ to $\mu^+ \mu^-$, $\tau^+ \tau^-$ and $\mu^{\pm}
\tau^{\mp}$.
 }}\label{B2ll}
\end{figure}
The partial decay width for the leptonic decay of $B_q$--meson
$\Gamma (\bar{B}_q \rightarrow \ell_i \ell_j )$ is given by
\begin{eqnarray}\label{Bbar->lilj}
\Gamma (\bar{B}_q \rightarrow \ell_i \ell_j ) = \frac{1}{16\pi
m_{B_q}^3} \overline{\vert \mathcal{A}\vert^2}
\lambda^{1/2}\left(m_{\ell_i}^2, m_{\ell_j}^2, m_{B_q}^2\right),
\end{eqnarray}
where $\overline{\vert \mathcal{A}\vert^2}$ is the spin averaged
amplitude for the radion contribution to the decay and
$$\lambda(x,y,z)=x^2+y^2+z^2 - 2xy - 2yz - 2zx~.$$

The experimental limits on the branching ratios are given as in
the table below

\begin{center}
\begin{tabular}{|c||c|}
  \hline
  Br & Experimental limit \\
  \hline\hline
  $B_s
\rightarrow \mu^+ \mu^-$ & $< 4.7\times 10^{-8}$ \\
$B_d
\rightarrow \mu^+ \mu^-$ & $< 1.5\times 10^{-8}$ \\
  $B_d \rightarrow \tau^+ \tau^-$ & $< 4.1\times 10^{-3}$ \\
  $B_d
\rightarrow \mu^+ \tau^-$ & $ < 2.2\times 10^{-5}$ \\
  \hline
  \hline
\end{tabular}
\end{center}

In Fig.\ref{B2ll} we display the branching ratio of $B_q$ decays
to $\mu^+ \mu^-$, $\tau^+ \tau^-$ and $\mu^{\pm} \tau^{\mp}$ as a
function of $m_\phi$ for $\Lambda_\phi =1$ TeV. From this figure,
one finds that the $Br(B_s \to \mu^+ \mu^-)$ can be of order
$10^{-8}$, \ie, within the range of accessibility at the LHCb, if
the radion mass is less than $10$ GeV. Also the present LHCb
experimental limit: $Br(B_s \to \mu^+ \mu^-) < 6.5 \times 10^{-8}
(95\%~{\rm C.L.})$ implies that $m_{\phi} > 6$ GeV. In addition,
it is predicted that within the region of light radion mass the
$Br(B_s \to \tau^+\tau^-)$ is of order $10^{-6}$, which can be
probed at the LHCb.

\section{Radion contribution to lepton flavor violating lepton decays} \label{sec:6}
In this Section we study lepton flavor violating (LFV) processes in which radion is either a decay product,
as in ${\ell}_i\longrightarrow{\ell}_j\bf{\phi}$, or which is mediated by a radion, {\it e.g.}
 $\ell_i \to \ell_j\ell_k\ell_l$.

\subsection{$\tau$ decay to a lepton and radion} \label{sec:6-1}
%
We start by studying lepton flavor violation via the process $\tau\to(e,\mu )\phi$.
We do not specify the decay products of the radion.
It dominantly decays to a gluon pair, but can decay also to a muon pair or a kaon pair, and with a small probability also to an electron pair or a pion pair.
Thus we have a muon or an electron from a tau decay, with no missing energy.
The limitation of this process is that we can only study radions which are lighter than $\tau$.

The amplitude for the decay is given by
\begin{eqnarray}\label{totalamplitude}
\mathcal{A}(\ell_i(k)\rightarrow\ell_j (p)\phi (q) 
&=&Y_{\phi\bar{\ell}_{Rj}\ell_{Li}}\bar{\ell}_{j}P_L\ell_{i} +Y_{\phi\bar{\ell}_{Lj}\ell_{Ri}}\bar{\ell}_{j}P_R\ell_{i}.
\end{eqnarray}
The total partial decay width for the LFV decay is
\begin{eqnarray}
\Gamma (\ell_i \to \ell_j \phi)) = \frac{1}{16\pi m_{\ell_i}^3} \overline{\vert \mathcal{A}\vert^2} \lambda^{1/2}\left(m_{\ell_i}^2,m_{\ell_j}^2,m_{\phi}^2\right)
\end{eqnarray}
where $M_\phi$ is the radion mass and $\lambda$ is defined after the equation (\ref{Bbar->lilj}). The spin-averaged amplitude in the rest frame of the decaying lepton is
\begin{eqnarray}
\overline{\vert \mathcal{A}\vert^2} =
 \left (Y_{\phi \bar{\ell}_{Lj}\ell_{Ri}}^2 + Y_{\phi\bar{\ell}_{Rj}\ell_{Li}}^2 \right)
\frac{1}{2} \left( m_{\ell_i}^2 + m_{\ell_j}^2 - M_{\phi}^2 \right)
+ 2 Re(Y_{\phi\bar{\ell}_{Lj}\ell_{Ri}}^*Y_{\phi\bar{\ell}_{Rj}\ell_{Li}}) m_{\ell_i} m_{\ell_j}.
\end{eqnarray}
%
The experimental value for $\Gamma(\tau\rightarrow (e,\mu)\phi)$ ,which can be calculated from the tau life time \cite{pdg1}, is found as
\begin{eqnarray}\label{allowed range}
\Gamma_{\tau}\simeq \left( 2.259692 \pm 0.00777 \right) \times
10^{-12}\,\texttt{GeV}.\,\,\,\,\,\,
\end{eqnarray}

\begin{figure}[!ht]
\centerline{
        \epsfxsize=3in
      \epsffile{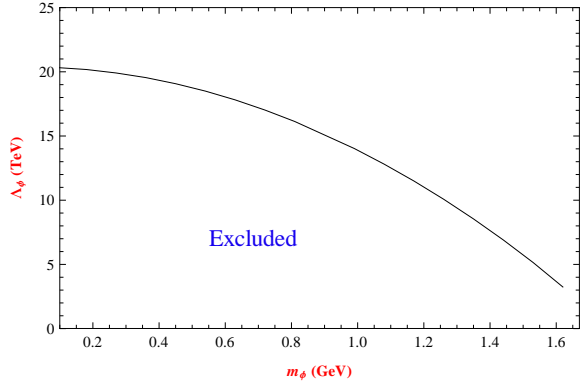}}
\caption{The excluded region in $(m_\phi,\Lambda_\phi)$ plane from
the $\tau\to(e,\mu)\phi$ decay at $2 \sigma$ level.} \label{taudecay}
\end{figure}

We have calculated the radion masses and $\Lambda_\phi$ which are allowed by the experimental error bars.
The result is plotted in Fig.~\ref{taudecay}.
We see that if the scale $\Lambda_\phi$ is large, this decay mode may still constrain the radion masses. For example, at $2 \sigma$ level if $\Lambda_\phi=10$ TeV, 
we found that the radion mass must be larger than 1.3 GeV. Thus, if we take $m_{\phi}\sim 1$ GeV we obtain $\Lambda_{\phi}\sim 15$ TeV.
We have checked that this value respects the condition Eq.~(12) in  Ref. \cite{Ponton}.
\newpage

\subsection{$\tau$ decay to three leptons} \label{sec:6-2}

\begin{figure}[!ht]
\centerline{
        \epsfxsize=1.8in
      \epsffile{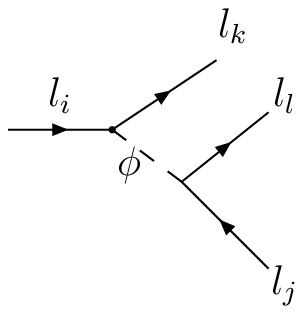}}
\caption{The decay  $\ell_i \to \ell_j\ell_k\ell_l$ through the
exchange of radion.}%
\label{muto3e}
\end{figure}

\begin{figure}[!ht]\centerline{
     \includegraphics[angle=270,width=100mm]{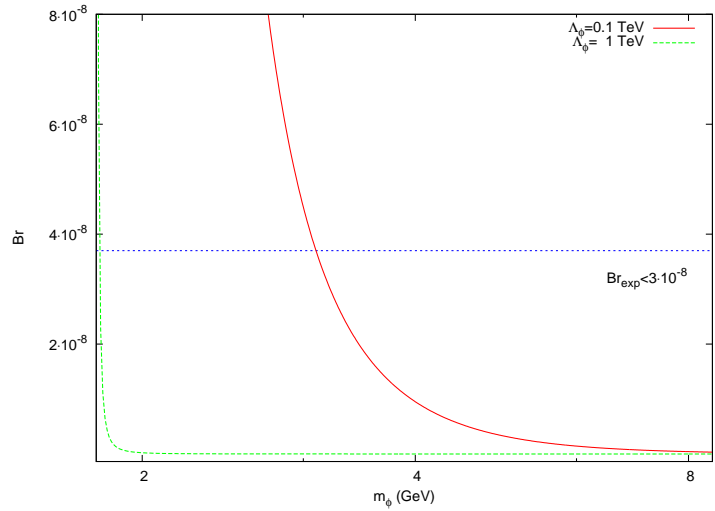}}
\caption{Branching ratio of $\tau\to e\mu\mu$ as a function of $m_\phi$ for the scale $\Lambda= 0.1$ and $1$ TeV.} \label{tau3ldecay}
\end{figure}

Much studied lepton flavor violating precision measurements
include $\ell_i\to\ell_j\ell_k\ell_l$ and $\ell_i\to\ell_j\gamma$.
As an example of this class of processes we study $\tau\to
e\mu\mu$, shown in Fig.~\ref{muto3e}, which has the largest
coupling constants in our parameter set for leptons.
%
%
%
%
%
The transition amplitude for the process $\tau\to e\phi^*\to e\mu\mu$ is given by%
\bea %
{\cal A}\Big(\tau(p)&\to& e^-(p_1) \mu^-(p_2)\mu^+(p_3)\Big)={1\over
q^2-m_\phi^2+i m_\phi \Gamma_\phi}\times \non\\
&&\sum_{\stackrel{i,j,k,l=L,R}{ i\neq j, k\neq l}}\Big[Y_{\phi \mu_i\mu^+_j}Y_{\phi\tau_k
e_l}(\bar{u}(p_2)P_jv(p_3))(\bar{u}(p_1)P_k u(p))\Big] .
\eea
%
Thus, one can show that the decay rate is given by
\begin{align}%
\label{gamma3e} \Gamma(\tau\to e\phi^*\to e\mu\mu)= {m_\tau^5\over
3\times 2^{12}\pi^3}\left({1\over m_\tau^2-m_\phi^2+i m_\phi \Gamma_\phi}\right)^2
\Big[Y_{\phi \mu_L \mu^+_R}^2Y_{\phi \tau_R e_L}^2+Y_{\phi \mu_L
\mu^+_R}^2Y_{\phi \tau_L e_R}^2  \nonumber \\ + Y_{\phi \mu_R
\mu^+_L}^2Y_{\phi \tau_R
e_L}^2+Y_{\phi \mu_R \mu^+_L}^2Y_{\phi \tau_L e_R}^2\Big].%
\end{align}
The experimental limit for the branching ratio is \cite{pdg1}:
\bea%
Br(\tau\to e\mu\mu)< 3.7\times 10^{-8}.%
\eea %
We show the branching ratio as a function of radion masses for $\Lambda_{\phi}=0.1$ and $1$ TeV in Fig.~\ref{tau3ldecay}. It is seen that this decay cannot compete with the other discussed ones in restricting $\Lambda_\phi$ vs $m_\phi$. With our parameter set, the same conclusion seems to hold for other $\ell_i\to$three leptons and also for lepton gamma modes.

%


%
\section{Radion Search at the LHC} \label{sec:7}
The radion coupling to the fermions is proportional to the fermion mass. The dominant production mode
for the radion at LHC is through gluon fusion.
The radion has an enhanced coupling with gluons through the trace anomaly \cite{Cheung:2000rw}: %
\be%
{\cal L}_{\rm int} = \frac{\phi(x)}{\Lambda_{\phi}}~ T_\mu^\mu,%
\ee%
with $T_\mu^\mu$ defined as %
\be%
T_\mu^\mu = \frac{\beta_{\rm QCD}}{2 g_s} {\rm
Tr}(F^{a}_{\mu\nu}F^{a\mu\nu})\ ,%
\ee%
where $F^a_{\mu\nu}$ is the field strength tensor of $SU(3)$
\begin{figure}[!ht]
\centerline{
        \epsfxsize=3.5in
      \epsffile{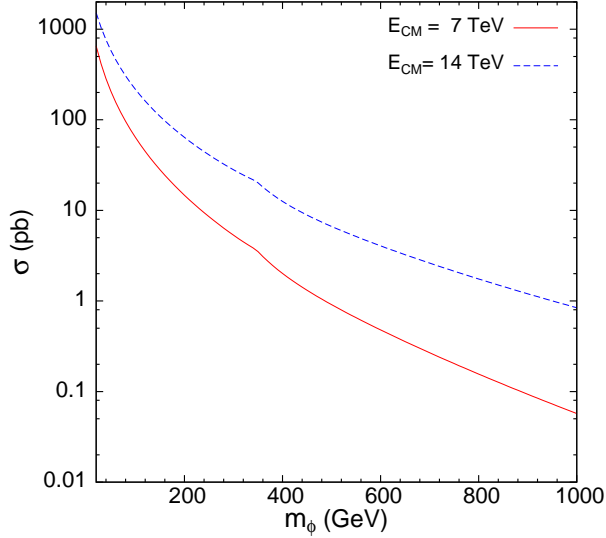}}
\caption{Radion production cross section at LHC for two different center of mass energies
as a function of the radion mass. We have fixed $\Lambda_\phi=1$ TeV. }%
\label{sigrad}
\end{figure}
interactions and $\beta_{QCD}$ is the QCD beta-function coefficient defined
as
\begin{eqnarray}
{\beta_{QCD}\over 2g_s}=-{\alpha_s\over 8 \pi}b_{QCD}.
\end{eqnarray}
Additional contributions coming from the heavy quark loop diagram is suppressed and neglecting that,
the cross section of radion production through gluon fusion at LHC can be written as
\begin{eqnarray}
\sigma_{LO}(gg\rightarrow \phi)=\int_{\tau}^{1} \frac{dx}{x}~G\left(x\right) G\left(\frac{\tau}{x}\right) ~ {\tau\alpha_s^2 \over
256\pi \Lambda^2_\phi }~|b_{QCD}|^2
\end{eqnarray}
where $b_{QCD}=11-2n_f/3$, $n_f$ being the number of quark flavors.  The gluon flux in the parton density functions (PDF) is given by $G(x)$
where $x$ is the momentum fraction carried by the gluons. For a radion of mass $m_\phi$ and the center of mass energy $\sqrt{s}$ we define
$\tau=m_\phi^2/s$. We calculate the total leading order cross section for the radion production at LHC for two different center-of mass energies, 7 TeV and
14 TeV as a function of the mass of the radion ($m_\phi$) with $\Lambda_\phi=1$ TeV and is shown in Fig. \ref{sigrad}. We use the {\tt Cteq6l}
PDF \cite{Pumplin:2002vw} for our calculation and the QCD scale $Q$ is set as the radion mass.
Note that the cross section scales as $1/\Lambda_\phi^2$. Thus if we reduce $\Lambda_\phi$ by a factor 2 then the cross section is increased
by a factor of 4. We have already shown that flavor physics constrains the parameter space with lower bounds obtained on the radion mass
for $\Lambda_\phi \sim {\cal O}(1)$ TeV. For a 100 GeV radion, $\Lambda_\phi$ can be as low as 300 GeV which implies a cross section of
$\sim$ 710 pb for the radion production at LHC with $\sqrt{s}=7$ TeV. For radion of mass less than 100 GeV, there are additional constraints on
$\Lambda_\phi$ from LEP data \cite{radionpheno}.

After its production, radion will decay either into $gg$, $W^+
W^-$, $ZZ$, $q_i \bar{q}_j$ or $\ell^+_i \ell^-$. Although the first
three channels dominate the radion decay \cite{Cheung:2000rw},
one can try to search for flavor violating radion decays through the
leptonic decay modes. The large cross section for the radion production
can give significant events for the alternative decay channels ($\ell^+_i \ell^-$) which
will have smaller SM background. Therefore, they could be striking
signatures for radion search at the LHC. However, the leptonic branchings of the radion are
very suppressed and fall rapidly with increasing radion mass. The flavor violating
leptonic decay channels ($\mu\tau/ e\tau$) are further suppressed with branching probabilities even smaller than
the diphoton channel and is of the order of $10^{-6}$ for light radion of mass less
than 100 GeV. The $\tau^+\tau^-$ decay channel is
about 5\% for a 50 GeV radion and $\Lambda_\phi=500$ GeV. With a good $\tau$-id at LHC,
this can be an important channel for the light radion signal at LHC. For a heavier radion with mass greater
than the top mass, the radion can decay to a top-quark and charm quark. This probability peaks
for a 250 GeV radion and has a branching probability of $\sim 1$\%. For values as low as
$\Lambda_\phi \sim 100$ GeV, this can give a $\sim$20\% contribution to the single top production which is
about $\sim 80$ pb in SM at LHC with $\sqrt{s}=7$ TeV. With the knowledge of the radion mass and with
dedicated cuts to isolate the signal from the background this mode can give hint to flavor violating decay
of the radion \cite{Azatov}. It will be however impossible to see any significant effects of
flavor violation in the leptonic sector at LHC in the ATLAS and CMS experiments from radion production. The heavy
radion would most likely be seen through its decays to the weak gauge bosons ($m_\phi > 140$ GeV ) while
the $\tau$ mode looks to be significant for the lighter radion. We refer the reader to various detailed studies
on radion signals at colliders \cite{radionpheno,radpheno}.

%
%

\section{Conclusions} \label{sec:8}
In this paper, we have analyzed the flavor violation in warped
extra dimension due to the radion exchange. In this scenario, the
SM fermions are propagating in the 5D bulk and the Higgs is
localized on the TeV brane. We found that $K -\bar{K}$ and
$B_q-\bar{B}_q$ lead to strong constraints on radion mass,
$m_\phi$ and the scale $\Lambda_\phi$. For instance, if
$\Lambda_\phi \sim {\cal O}(1)$ TeV, one finds that $B_d^0
-\bar{B}^0_d$ implies that $m_\phi \gsim 65$ GeV. We have also
studied the radion contributions to lepton flavor violating
processes: $\ell_i \to \ell_j \phi$ and $\ell_i \to \ell_j \ell_k
\ell_l$, in addition to $B\to \ell_i \ell_j$. We have shown that
the $BR(\tau \to (e,\mu) \phi)$ imposes a stringent limit on the
scale $\Lambda_\phi$ for $m_\phi \lsim {\cal O}(1)$ GeV. We
emphasized that the radion effect to $BR(B_s \to \mu^+ \mu^-)$ can
be of order $10^{-8}$, which is accessible at the LHCb. We have
also analyzed the search for radion at LHC. Although, we do not
find any significant flavor violating signals in the lepton
sector, there is definitely a possibility of contributions to
single top cross section with the radion decaying through the
flavor violating mode of $t\bar{c}+\bar{t}c$.

\section*{Acknowledgments}
KH and AS acknowledge support from the Academy of Finland (Project No.
137960) and AS also from the Finnish Cultural Foundation and CIMO.
The work of S.K. and A.M. was partially supported by the Science
and Technology Development Fund (STDF) project ID 1855 and the
ICTP project ID 30. SKR is supported in part
by the US Department of Energy, Grant Number DE-
FG02-04ER41306.

\bibliographystyle{unsrt}

\end{document}